\documentstyle[amssymb,twocolumn,prl,aps]{revtex}
\begin{document}
\twocolumn[
\hsize\textwidth\columnwidth\hsize\csname@twocolumnfalse\endcsname
\draft
\title { Optical properties of an effective one-band Hubbard model
for the cuprates} 
\author{ M.E.Sim\'on, A.A.Aligia and E.R.Gagliano}
\address{Centro At\'omico Bariloche and Instituto Balseiro}
\address{Comisi\'on Nacional de Energ{\'\i}a At\'omica}
\address{8400 S.C. de Bariloche, Argentina.}

\date{Received \today}
\maketitle
  
\begin{abstract}
We study the Cu and O spectral density of states and the optical
conductivity of CuO$_2$ planes using an effective generalized one-band
Hubbard model derived from the extended three-band Hubbard model. 
We solve exactly a square cluster of 10 unit cells and average the
results over all possible boundary conditions, what leads to smooth
functions of frequency.
Upon doping, the Fermi energy jumps to Zhang-Rice states which are
connected to the rest of the valence band ( in contrast to an isolated new 
band in the middle of the gap). The transfer of spectral weight depends
on the parameters of the original three-band model not only through the
one-band effective parameters but also through the relevant matrix elements.
We discuss the evolution of the gap upon doping. The optical conductivity of
the doped system shows a mid-infrared peak due to intraband transitions, a
pseudogap and a high frequency part related to interband transitions.
Its shape and integrated weight up to a given frequency ( including the 
Drude weight ) agree qualitatively with experiments in the cuprates for 
low to moderate doping levels, but significant deviations exist for 
doping $x>0.3$.

\end{abstract}
]

\narrowtext

\section{\bf INTRODUCTION}

In recent years, there has been much interest in the spectral and related
electronic properties of cuprates superconductors and similar materials \cite
{dess,sree,uchi,falck,nuck,chen,hybe,romb,alle,shen,perk,ayac,tera,fuji,roze,mill,wagn,esk1,sf,che2,dag1,dag2,ohta,poil,lore,hors,bulu,jakl,eder,eske,unge}%
. In spite of the considerable research effort, there are several issues
which remain to be clarified. For example, photoemission experiments at
optimal doping\cite{dess} show that electrons have a large Fermi surface of
area $(1-x)$ where $x$ is the amount of hole doping, while at small doping
it is expected that the Fermi surface consists of four small hole pockets
centered at $(\pm \pi/2,\pm \pi/2)$ of total area $x$. The evolution of the
Fermi surface remains a tough problem \cite{unge}. There is a recent
theoretical study on this subject\cite{chub}.

Another related issue is the appearance and evolution for $x\neq 0$ of states at
energies which lie in the gap for $x=0$. There are at least two physical
pictures. On the basis of several measurements, and particularly the change
of sign of the Hall constant with $x$ in La$_{2-x}$Sr$_x$CuO$_4$, Sreedhar
and Ganguly \cite{sree} proposed that the gap in the spectral density
gradually closes with increasing $x$, and after the closing, the system has
completed its evolution from a Mott ( or better said "charge-transfer" )
insulator to a band metal. As soon as doping begins the Fermi level jumps to
the valence band. This jump is consistent with the measured x-ray
absorption spectrum (XAS)\cite{chen,hybe}. Instead, from the optical
conductivity\cite{uchi} and high-energy spectroscopies\cite{nuck}, it has
been suggested that a new band of midgap states is formed, taking spectral
weight from both, conduction and valence bands, and the Fermi level remains
in the midgap band. A minimal shift of the Fermi level with doping has been
observed by photoemission in La$_{2-x}$Sr$_x$CuO$_4$\cite{romb} and in the
electron doped Nd$_{2-x}$Ce$_x$CuO$_4$\cite{alle}, but not in Ba$_2$Sr$_2$Ca$%
_{1-x}$Y$_x$CuO$_2$C$_8$\cite{shen}. It might be possible that the states in
the middle of the gap are created by deep donor levels originated by the
substitution of La$^{+3}$ by Sr$^{+2}$ ions. The resulting "impurity-state
model" is consistent with the experimental evidence in the oxide (Nd,Sr)CoO$%
_3$ \cite{tera}. The effect of this substitution has been discussed in Ref.%
\cite{hybe}. However, inhomogeneities are usually not included in the
theoretical treatments, and it seems necessary to generalize the Hubbard
model to explain the electronic structure of some non-stoichiometric oxides%
\cite{tera,fuji}.

To discuss the validity of the translationally invariant three-band Hubbard
model\cite{3b} or effective models derived from it \cite
{sf,che2,fei1,sim1,erol} in the description of the cuprates, it is necessary
to know precisely the spectral properties of these models. One of the most
used and effective ways to study these properties in two dimensions at zero
temperature is the exact diagonalization of small clusters\cite
{wagn,sf,che2,dag1,dag2,poil,eder}. Among the different effective models,
those similar to the $t-J$ one \cite{dag2,poil,eder,fei1,erol} have the
smallest number of states per unit cell. However, like the spin-fermion ( or
Kondo-Heisenberg) model\cite{sf,che2}, their Hilbert space is too small to
allow a description of both, valence and conduction bands. In contrast, the
number of states per unit cell of the three-band model is so large that the
largest exactly solved cluster contains only four unit-cells\cite
{wagn,esk1,ohta}. Thus, the most apropriate effective model to study the
evolution of the gap seems to be the extended Hubbard one\cite{sim1}.
Numerical studies of the Hubbard model show a change of sign of the Hall
constant as a function of doping, but it has not been related with a
reconstruction of the spectral density\cite{cast,shastry,rojo,assa}. The
spectral properties of the Hubbard model have been calculated in periodic
square clusters of 8,10 and 16 sites \cite{dag1,dag2,ohta}. Without
artificial broadening, the result of these calculations is given by a set of
delta functions at different frequencies ( see for example Fig.11 of Ref\cite
{dag2}), from which it is not possible to distinguish if in the
thermodynamic limit there is one, two or more gaps in the spectral density
of states. Also the Drude weight shows important finite-size effects, being
negative near half-filling.

In this work we solve exactly the effective one-band extended Hubbard
hamiltonian in a square cluster of 10 unit cells and average over all
possible boundary conditions\cite{gros}. This technique has been also used
by Poilblanc to study the optical conductivity of the $t-J$ model\cite{poil}
and by Xiang and Wheatly \cite{xiang} to calculate the dispersion relation
of one hole in a generalized $t-J$ model, obtaining good agreement with
results of the self-consistent Born approximation and experiments in Sr$_2$%
CuO$_2$Cl$_2$. To our knowledge this approach has not been applied to the
spectral properties of Hubbard or extended Hubbard clusters. It allows us to
obtain continuous spectral densities and optical conductivities and
according to our studies in small rings ( see section III ), the convergence
to the thermodynamic limit is faster.

It must also be pointed out that the spectral properties of the Hubbard
model derived from the numerical methods mentioned above, as well as
analytical studies of the change in the spectral weights with doping\cite
{eske} {\it do not} directly correspond to the cuprates as described by the
original three-band Hubbard model. To calculate any property of this model
using an effective Hamiltonian, one should transform the corresponding
operators from the three-band to the effective Hamiltonian. This has been
done for the spin-fermion\cite{sf}, Hubbard\cite{fei2} and $%
t-t^{\prime}-t^{\prime\prime}-J$\cite{erol} as effective models for the
cuprates. In particular, Feiner explained the electron-hole asymmetry in the
spectral weight in the cuprates ( which is absent in the "isolated" Hubbard
model)\cite{fei2}. Here we use and discuss the transformed operators for
spectral density and optical conductivity.

In section II, we briefly review the extended one-band Hubbard model as an
effective model for cuprate superconductors, and construct the relevant
operators. In section III, we discuss the numerical techniques and the
method of integrating over arbitrary boundary conditions. In section IV, we
present the results for the spectral density, while section V contains the
results for the optical conductivity, Drude weight and related quantities.
The conclusions are presented in Section VI.

\section{\bf The effective extended one-band Hubbard model and transformed
operators}

Experimental evidence about the symmetry of holes in high-$T_c$
superconductors\cite{nuck,taki,pele}, as well as
constrained-density-functional calculations\cite{anne,hyb2}, support the
appropriateness of the three-band Hubbard model \cite{3b} for the
description of these systems. The Hamiltonian is: 
\begin{eqnarray}
H_{3b} = \Delta \sum_{j\sigma}p^{\dagger}_{j\sigma}p_{j\sigma} +
t_{pd}\sum_{i\delta\sigma}(p^{\dagger}_{i+\delta\sigma}d_{i\sigma}+h.c.)- \\
t_{pp}\sum_{j\gamma\sigma}p^{\dagger}_{j+\gamma\sigma}p_{j\sigma} +
U_{pd}\sum_{i\delta\sigma\sigma^{\prime}}
n^{d}_{i\sigma}n^{p}_{i+\delta\sigma^{\prime}}+  \nonumber \\
U_{d}\sum_{i} n^{d}_{i\uparrow} n^{d}_{i\downarrow} + U_{p}\sum_{j}
n^{p}_{j\uparrow} n^{p}_{j\downarrow}  \nonumber \\
\nonumber
\end{eqnarray}

\noindent where $d^{\dagger}_{i\sigma}$ ($p^{\dagger}_{j\sigma}$) creates a
hole in the $d_{x^{2}-y^{2}}$ ( $p_\sigma$ ) orbital at site $i$ ( $j$ )
with spin $\sigma$, $i+\delta$ ( $j+\gamma$ ) label the four O atoms
nearest-neighbors to Cu (O) site $i$ ( $j$ ). The phases of half the $d$ and 
$p$ orbitals have been modified by a factor (-1) in order to have $t_{pd}>0$
and $t_{pp}>0$ independently of the direction. While other Cu and O orbitals
should be included to explain Raman experiments\cite{liu,rap}, these
additional states affect neither low-energy spectral properties 
(below the bottom of the $p_\pi$ band \cite{rap} ) nor
the validity of the one-band effective model $H_{1b}$ \cite{rap}.

The transformation of the low-energy part of $H_{3b}$ to the effective model 
$H_{1b}$ can be summarized in four steps\cite{fei1,sim1,rap,schu}: (i)
change of basis of the O orbitals $p_{j}^{x},p_{j}^{y}$ to orthogonal
Wannier functions $\alpha_{i},\gamma_{i}$ centered at each Cu site $i$, with
symmetries $b_{1g}$ ( the same as the $d_{x^{2}-y^{2}}$ orbital ) and $a_{1g}
$ respectively; (ii) exact solution of the cell Hamiltonian $H_i$ ( the
terms of $H_{3b}$ which contain only operators acting on cell $i$ ); (iii)
mapping the low-energy states $|m\rangle$ of $H_{3b}$ into corresponding
ones $|\bar{m}\rangle$ of the Hilbert space of the one band model ($%
|m\rangle \leftrightarrow |\bar{m}\rangle$ ); (iv) transform the operators $%
H_i$ and $H_{3b}-\sum_{i}H_{i}$ to the new basis $|\bar{m}\rangle$. A
transformed operator $\bar{O}$ can be expressed in terms of the matrix
elements of the original operator $O$ as: 
\begin{eqnarray}
\bar{O} = \sum_{n,m} \langle n|O|m \rangle |\bar{n} \rangle \langle \bar{m}|
\end{eqnarray}

\noindent Finally, the effect of states of $H_{3b}$ which do not have
corresponding ones in $H_{1b}$ can be included as perturbative corrections 
\cite{sim1}, but we will not do it here. In Fourier space, the $p_\sigma$
orbitals are related with the O Wannier functions by: 
\begin{eqnarray}
p^{x}_{{\bf k}\sigma } = \beta_{{\bf k}} ( \cos {\ ( {\frac{ {\ k_{x}a}}{2 }}
) } \alpha_{{\bf k}\sigma } - \cos {\ ( {\frac{ {\ k_{y}a}}{2 }} ) } {\
\gamma_{{\bf k}\sigma } }) \\
p^{y}_{{\bf k}\sigma } = \beta_{{\bf k}} ( \cos {\ ( {\frac{ {\ k_{y}a}}{2 }}
) } \alpha_{{\bf k}\sigma } + \cos {\ ( {\frac{ {\ k_{x}a}}{2 }} ) } {\
\gamma_{{\bf k}\sigma } } )  \nonumber
\end{eqnarray}

\noindent with $\beta_{{\bf k}}= {(1 + {\frac{1}{2}} \cos{\ ( k_{x}a ) } + {%
\frac{1}{2}} \cos{\ ( k_{y}a ) })}^{-{\frac{1}{2}}}$. The eigenstates of $H_i
$ retained in the mapping procedure are ( in addition to the vacuum at site $%
i$ ), the lowest one-hole doublet $|i\sigma\rangle$ and the lowest two-hole
doublet $|i2\rangle$. They can be expressed as: 
\begin{eqnarray}
|i2\rangle = [{\frac{ {A_{1}}}{\sqrt{2} }} ( d^{\dagger}_{i\uparrow}
\alpha^{\dagger}_{i\downarrow} - d^{\dagger}_{i\downarrow}
\alpha^{\dagger}_{i\uparrow} ) - A_{2}
\alpha^{\dagger}_{i\uparrow}\alpha^{\dagger}_{i\downarrow} - A_{3}
d^{\dagger}_{i\uparrow}d^{\dagger}_{i\downarrow} ]|0\rangle \\
|i\sigma\rangle = [B_{1}d^{\dagger}_{i\sigma} -
B_{2}\alpha^{\dagger}_{i\sigma}] |0\rangle  \nonumber
\end{eqnarray}

\noindent These states are mapped into those of the one-band model by the
correspondence $|i\sigma \rangle \leftrightarrow c_{i\sigma }^{\dagger
}|0\rangle $, $|i2\rangle \leftrightarrow c_{i\uparrow }^{\dagger
}c_{i\downarrow }^{\dagger }|0\rangle $. This assignment and Eq.(2) applied
to $H_{3b}$ leads to the effective one-band generalized Hubbard Hamiltonian 
\cite{sim1}, with occupation dependent nearest-neighbor hopping and
repulsion ( interactions and hoppings at larger distances are neglected for
simplicity ): 
\begin{eqnarray}
H_{1b} &=&E_1\sum_{i\sigma }n_{i\sigma }+U\sum_in_{i\uparrow }n_{i\downarrow
}+\sum_{<ij>\sigma }c_{j\sigma }^{\dagger }c_{i\sigma }\{t_{AA}(1-n_{i,{\bar{
\sigma}}})(1-n_{j,{\bar{\sigma}}})+ 
t_{AB}[n_{i,{\bar{\sigma}}}(1-n_{j,{\bar{\sigma}}})+n_{j,{\bar{\sigma}}
}(1-n_{i,{\bar{\sigma}}})]+ \\ \nonumber
&&t_{BB}n_{i,{\bar{\sigma}}}n_{j,{\bar{\sigma}}
}\}+h.c.+ 
\sum_{<ij>\sigma \sigma ^{\prime }}\{V_{11}(1-n_{i,{\bar{\sigma}}})(1-n_{j,
{\bar{\sigma}}^{\prime }})+V_{22}n_{i,{
\bar{\sigma}}}n_{j,{\bar{\sigma}}^{\prime }}+ \\ \nonumber
&&V_{12}[(1-n_{i,{\bar{\sigma}}})n_{j,{
\bar{\sigma}}^{\prime }}+n_{i,{\bar{\sigma}}}(1-n_{j,{\bar{\sigma}}^{\prime
}})]\}n_{i,\sigma }n_{j,\sigma ^{\prime }}  \nonumber
\end{eqnarray}

\noindent Several particular cases of this model have been widely studied 
\cite{hirs,japa,arr1,mich,arr2,ali1,oht2}. For $t_{AA}+t_{BB}=2t_{AB}$ and $%
V_{ij}=0$, the model is shown to lead to superconductivity for small $U$ and
adequate doping\cite{hirs}, particularly in one dimension \cite
{japa,arr1,mich}. For $V_{ij}=t_{AB}=0$, the model has been exactly solved
in one dimension \cite{arr2}. For $t_{AB}=0$, $V_{22}=V_{21}=V_{11}$ and
half-filling, the exact ground state has been found in a wide range of
parameters for arbitrary lattices in arbitrary dimension, and the phase
diagram separating metallic, Mott insulating and charge-density waves
regions has been established\cite{ali1}.

The spectral properties in two dimensions have been studied numerically in
the charge-density-wave ( large $V_{ij}$ ) regime \cite{oht2}, and for $%
V_{ij}=0$ and small $t_{AB}$ at half-filling \cite{gag1}. The model has been
also used recently to study the regions of phase separation and valence
instabilities of the original three-band model\cite{sim1}. Typical values of
the parameters of $H_{3b}$ and $H_{1b}$ are given in tables I and II
respectively.

At zero temperature, the Cu and O spectral densities are defined by: 
\begin{eqnarray}
\rho_{O}(\omega) = {\frac{1}{L}} \sum_{k,n} |\langle n|\bar{O}_{k}|g\rangle
|^{2} \delta(\omega+E_{g}-E_{n})
\end{eqnarray}

\noindent where $|n\rangle$,$E_{n}$ are the eigenstates and energies of $%
H_{1b}$, $|g\rangle$ labels the ground state for a given number of particles 
$N$ and $L$ is the number of unit cells in the system.
For electron spectral densities of
Cu or one of both O atoms in the unit cell ( measured by inverse
photoemission ), $\bar{O}_k$ is the operator in the representation of $H_{1b}
$ of $d_{k}$ or $p_{k}^{x,y}$ respectively. For hole spectral densities (
corresponding to photoemission ) the corresponding hermitian conjugate
operators should be taken. In real space using Eq.(2) one obtains: 
\begin{eqnarray}
\bar{d}_{i} & = & D_{1} c_{i\sigma} (1-n_{i\bar\sigma}) + D_{2} c_{i\sigma}
n_{i\bar\sigma} \\
\bar{\alpha_{i}} & = & P_{1} c_{i\sigma} (1-n_{i\bar\sigma}) + P_{2} c_{i\sigma}
n_{i\bar\sigma}  \nonumber
\end{eqnarray}

\noindent where 
\begin{eqnarray}
D_{1}& = & \langle i0|d_{i\sigma}|i\sigma\rangle = B_{1} \\
D_{2}& = & \langle i{\bar\sigma}|d_{i\sigma}|i2\rangle = ( {\frac{ {%
A_{1}B_{2}}}{\sqrt{2} }} + A_{3}B_{1} )  \nonumber \\
P_{1}& = & \langle i0|\alpha_{i\sigma}|i\sigma\rangle = -B_{2}  \nonumber \\
P_{2}& = & \langle i{\bar\sigma}|\alpha_{i\sigma}|i2\rangle = -( {\frac{ {%
A_{1}B_{1}}}{\sqrt{2} }} + A_{2}B_{2} )  \nonumber
\end{eqnarray}


\vskip 4 truecm


\begin{center}
\end{center}

\noindent Transforming Fourier and using Eq.(3), the operators entering
Eq.(6) are defined. As will be discussed in section IV, the asymmetry of
Eq.(7) and also the Hamiltonian Eq.(5) under electron-hole transformation
leads to a different behavior of the spectral densities under electron or
hole doping. This difference is apparent in the cuprates and in the
three-band Hubbard model results, but it is absent in the ordinary one-band
Hubbard model \cite{esk1,fei2}.

The optical conductivity can be written as \cite{mill,wagn} 
\begin{eqnarray}
\sigma(\omega) = 2\pi D \delta(\omega) + \sigma_{reg}(\omega)
\end{eqnarray}

\noindent where 
\begin{eqnarray}
\sigma_{reg}(\omega) = {\frac{\pi }{{V\omega} }} \sum_{n \neq g} |\langle
n|j_{x}|g \rangle |^{2} \delta(E_{n}-E_{g}-\omega)
\end{eqnarray}

\noindent $V$ is the volume of the system, $j_x$ the $x$ component of the
current operator. The Drude weight $D$ can be calculated in two ways: either
using the mean value of the kinetic energy in a given direction $\langle
T_{x} \rangle$: 
\begin{eqnarray}
D= {\ ({\frac{ {ea}}{\hbar }})^{2} } {\frac{ {\langle -T_{x}\rangle}}{{2V} }}
- {\frac{1}{V}}\sum_{n \neq g} {\frac{ {|\langle n|j_{x}|g \rangle |^{2}}}{{%
E_{n}-E_{g}} }}
\end{eqnarray}

\noindent or from the second derivative of the energy with respect to the
vector potential $A$ in a system with periodic ( or arbitrary as described
in the next section ) boundary conditions 
\begin{eqnarray}
D = {\frac{ {\ c^{2} } }{{2V}}} {\frac{ {d^{2} E(A)}}{{dA^{2}} }} {\Bigl|}%
_{A=0}
\end{eqnarray}

\noindent The current operator can be derived in a standard way, from $dH/dA$%
, when each hopping term $c^{\dagger}_{i\sigma}c_{j\sigma}$ is multiplied by
a phase factor Exp$[ieA(x_{i}-x_{j})/(c\hbar)]$ \cite{mill,wagn}. In the
case of the effective Hamiltonian Eq.(5), the current can be divided in
three terms proportional to the three different types of hopping terms: 
\begin{eqnarray}
j_{x} = j_{AA} + j_{AB} + j_{BB}
\end{eqnarray}
\noindent
with 
\begin{eqnarray}
j_{AA}(r) &=& {\frac{ {iae}}{\hbar }} t_{AA} \sum_{\sigma}
(c^{\dagger}_{r\sigma}c_{r+x\sigma}-c^{\dagger}_{r+x\sigma}c_{r\sigma})
(1-n_{r,{\bar \sigma}}) (1-n_{r+x,{\bar \sigma}}) \\
j_{AB}(r) &=& {\frac{ {iae}}{\hbar }} t_{AB} \sum_{\sigma}
(c^{\dagger}_{r\sigma}c_{r+x\sigma}-c^{\dagger}_{r+x\sigma}c_{r\sigma})
[(1-n_{r,{\bar \sigma}})n_{r+x,{\bar \sigma}} + n_{r,{\bar \sigma}}(1-n_{r+x,%
{\bar \sigma}}) ]  \nonumber \\
j_{BB}(r)& = &{\frac{ {iae}}{\hbar }} t_{BB} \sum_{\sigma}
(c^{\dagger}_{r\sigma}c_{r+x\sigma}-c^{\dagger}_{r+x\sigma}c_{r\sigma}) n_{r,%
{\bar \sigma}}n_{r+x,{\bar \sigma}}  \nonumber
\end{eqnarray}

An alternative derivation of the current operator for the effective one-band
model is to obtain first the current operator of the multiband model\cite
{wagn}, then project the result into the low-energy subspace and map it onto
the one-band representation. Both derivations lead to identical results if $%
t_{pp}=0$. For realistic values of $t_{pp}$, there are small differences
between the resulting coefficients entering the current $j_{AA},~~j_{AB}$ 
and $j_{BB}$. For example while for the set 2 of parameters of $H_{3b}$ (see
table I ) we obtain $t_{AA}=t_{AB}=-0.38$, $t_{BB}=-0.33$, in the second
derivation these values should be replaced by -0.32, -0.35 and -0.36 in the
respective currents. This difference is probably due to the different way in
which excited ( mainly local triplet ) states are taken into account in
lowest order in both procedures when $t_{pp}\neq 0$\cite{trip,erol}. In this
work we use Eq.(14).

\section{\bf Numerical methods and average over boundary conditions}

We have evaluated Eqs.(6), (10) and (11) using the now standard
continued-fraction expansion of these equations with the Lanczos method \cite
{gag2}, in a square cluster of 10 unit cells, integrating the result over
the different boundary conditions\cite{gros,poil}. Particular boundary
conditions are specified by two phases $(\phi_{1},\phi_{2})$, $0\leq
\phi_{i}<2\pi$ in the following way: an infinite square lattice is divided
into square 10-site clusters, and the sites which are at distances $n{\bf %
V_{1}}+m{\bf V_{2}}$ with $n,m$ integers and ${\bf V_{1}}=(3,1),~~{\bf V_{2}}%
=(-1,3)$ are considered equivalent ( see Fig.1). After choosing a particular
10-site cluster, each "time" the hopping term makes a particular jump out of
the cluster, it is mapped back into the cluster through a translation in one
of the vectors ${\bf -V_{1}},{\bf V_{1}},{\bf -V_{2}},$ or ${\bf V_{2}}$ and
the wave function is multiplied by $e^{i\phi_{1}}, e^{-i\phi_{1}},
e^{i\phi_{2}}$ or $e^{-i\phi_{2}}$ respectively. When $(\phi_{1},%
\phi_{2})=(0,0)$ this is equivalent to the usual periodic boundary
conditions. It is easy to see that when we impose the wave function for $N$
particles to be an irreducible representation of the group of translations
of the infinite square lattice, the allowed total wave vectors ${\bf K}$
should satisfy 
\begin{eqnarray}
N\phi_{1} + 2n_{1}\pi = {\bf K} \cdot {\bf V_{1}} \\
N\phi_{2} + 2n_{2}\pi = {\bf K}\cdot {\bf V_{2}}  \nonumber
\end{eqnarray}


\vskip 4 truecm


\begin{center}
\end{center}

\noindent where $n_{1}, n_{2}$ are integers. Solving for $K_{x},K_{y}$ one
obtains: 
\begin{eqnarray}
K_{x} = {\frac{ N}{10 }} (3\phi_{1}-\phi_{2} ) - {\frac{ {2\pi(n_{2}-3n_{1})}%
}{10}} \\
K_{y} = {\frac{ N}{10 }} (\phi_{1}+3\phi_{2} ) + {\frac{ {2\pi(3n_{2}+n_{1})}%
}{10}}  \nonumber
\end{eqnarray}
\noindent When $(\phi_{1},\phi_{2})=(0,0)$, varying the integers $n_{1},
n_{2}$, Eqs.(16) give the 10 inequivalent wave vectors allowed by periodic
boundary conditions. When $(\phi_{1},\phi_{2})$ is allowed to vary
continuously the whole reciprocal space can be swept. This allows us to
obtain more information from the finite-size cluster calculations. Also,
while the spectral densities and optical conductivities being calculated are
given by a sum of delta functions for each $(\phi_{1},\phi_{2})$ ( Eqs.(6)
and (9)), integrating the result over $(\phi_{1},\phi_{2})$ leads to
continuous functions. In practice we have replaced the integral by an
average over a square mesh of up to 8x8 points ( up to 34 non-equivalent by
symmetry for the optical conductivity) in $\phi_{1},\phi_{2}$ space until a
fairly smooth function was obtained.

However, the procedure described above has some shortcomings when applied to
the optical conductivity. For general boundary conditions there is a {\it %
spontaneous current} in the ground state, invalidating the derivation of
Eqs.(9) to (12)\cite{mill,wagn}. (Such a current can not exist in the
thermodynamic limit, even if it were allowed by symmetry \cite{blou,sim3}).
In finite systems, this current is zero for periodic boundary conditions at
"closed-shell" fillings, (including half-filling in our cluster), and for
half-filling and any boundary conditions in the $U\rightarrow \infty$ limit
(as in the $t-J$ model \cite{poil} ), since in this limit the charge
dynamics is suppressed. Our point of view is that the average of Eqs.(10)
and (11) over boundary conditions in a finite system, is an approximation to
the correct expression in the thermodynamic limit, which converges faster
with system size than the result of Eqs.(10) and (11) or (12) for fixed
 boundary conditions. Since we have not done finite-size scaling in
two dimensions, we can not prove this statement. However, it is supported by
our one-dimensional check summarized below. Furthermore, as we shall see in
section V, the results using the averaging procedure look reasonable, while
the Drude weight obtained in larger systems using periodic boundary
conditions show unphysical negative values at half-filling \cite{dag2,bulu}.
Previous calculations of the optical conductivity of the $t-J$ model using
twisted boundary conditions also support our procedure\cite{poil}.

We have checked the method of averaging over boundary conditions applying it
to the optical conductivity $\sigma(\omega)$ of one-dimensional Hubbard
rings, for which finite-size scaling can be done, and some exact results,
like the charge gap, are available\cite{ovch}. The results for $U/t=4,6$ and 
$8$ and several system sizes $L$ are represented in Figs.2-4 respectively. A
funny feature of these curves is the presence of oscillations in $%
\sigma(\omega)$. A comparison of the result for rings of different sizes
shows that this is a finite-size effect. Calculations using spin-wave theory
in the strong coupling limit give a smooth $\sigma(\omega)$ \cite{hors}.
Fortunately no signs of these oscillations are present in the
two-dimensional case. By comparison, the strong coupling limit of the
Hubbard model, and the $t-J$ model with one hole in systems as large as 19
sites with periodic boundary conditions shows a few delta functions\cite
{hors}.

The presence of a peak at zero frequency is a consequence of the already
discussed spontaneous current in the ground state for general boundary
conditions. Clearly, this is also a finite-size effect and the height of the
peak decreases with system size and also with increasing $U/t$. Thus, we
have suppressed this peak in the two-dimensional results of section V.

We have also noticed that the gap in $\sigma(\omega)$ converges much faster
to the thermodynamic limit $E_{g}$ than the corresponding result using fixed
boundary conditions. In table III, we compare the exact gap, given by the
integral\cite{ovch};
\begin{eqnarray}
{\frac{E_{g}}{t}} = {\frac{ {16t}}{U }} \int_{1}^{\infty} {\frac{ {\sqrt{%
y^{2}-1}}}{{\sinh( {\frac{{2\pi t y }}{U }} )} }} dy
\end{eqnarray}

\noindent against its value ( taken as the position of the lowest frequency 
peak in the spectra of Figs. 2-4 ) for a ring of 8 sites after averaging over
boundary conditions and at fixed antiperiodic boundary condition.
Clearly, the averaging procedure speeds up the convergence toward the
$L\infty$ limit.
 
\section{\bf The spectral densities}

In Fig.5, we show the Cu and O spectral densities $\rho(\omega)$ in the hole
representation ( photoemission and inverse photoemission correspond to $%
\omega >\epsilon_{F}$ and $\omega < \epsilon_{F}$ respectively, where $%
\epsilon_F$ is the Fermi energy ) for several doping levels and the set of
parameters 1 ( see Tables I and II ). These spectral densities have been
calculated as described in section III, using Eq.(6) with $\bar{O}_k$
replaced by $\bar{d}^{\dagger}_k$, $\bar{p}^{x\dagger}_k$ for Cu and O
photoemission, and by $d_k$, $p^{x}_{k}$ for inverse photoemission. The
result for O has been multiplied by a factor 2 to represent the total
contribution of both $2p_\sigma$ orbitals of the unit cell. Also shown in
Fig.5 are the extended Hubbard spectral densities corresponding to the
operators $c^{\dagger}_{k}(\omega > \epsilon_{F})$ and $c_{k}(\omega <
\epsilon_{F})$. Note that in constrast to the usual Hubbard model, the
extended Hubbard spectral densities are not symmetric under hole $(x>0)$ or
electron $(x<0)$ doping. This is due mainly to the fact that $t_{AA}\neq
t_{BB}$. However, the asymmetry of the Cu and O spectral densities is much
stronger. In agreement with experiment, for hole doping, the inverse
photoemission spectrum near $\epsilon_F$ shows a larger contribution of O
states than Cu $d$ states, while for electron doping, the photoemission
spectrum near the Fermi level is dominated by Cu states.

A pseudogap in the spectral densities persists with doping and allows to
separate the electronic states in two bands at lower and higher energies
than this pseudogap. The lower band corresponds to states with low double
occupancy and therefore, the spectral densities are dominated by the
contribution proportional to $c_{i\sigma}(1-n_{i\bar{\sigma}})$ in Eq.(7);
thus the Cu and O spectral densities in the lower band are, as a first
approximation proportional to the respective coefficients( $B_{1}$ and $B_{2}
$, see Eq.(8) ) in the one-particle ground state of the cell. However, the
states of the lower band contain some admixture of double occupied states,
and this introduces an energy dependence of the relative weight of Cu and O
states. The states of lower energy $(\omega \sim -1)$ have a larger Cu
content than those of the rest of the lower band. Similarly, as discussed
previously by Feiner\cite{fei2}, the Cu and O contents of the upper band are
mainly determined by $D_2$ and $P_2$ respectively ( see Eq.(8)), which
depend on the structure of the Zhang-Rice singlet $|i2\rangle$ $and$ the
ground state of the cell with one hole $|i\sigma\rangle$ (see Eq.(4) ).

From Eq.(6) applied to the Hubbard operators $c_k$ and $c^{\dagger}_k$, it
is easy to obtain the total inverse photoemission $N^-$ and photoemission $%
N^+$ spectral weight: 
\begin{eqnarray}
N^{-}&=&\int_{-\infty}^{\epsilon_F} \rho_{c}(\omega)d\omega =
<n_{i\uparrow}>+ <n_{i\downarrow}> = 1+x \\
N^{+}&=&\int_{\epsilon_F}^{\infty} \rho_{c^{+}}(\omega)d\omega = 2 -
<n_{i\uparrow}>- <n_{i\downarrow}> = 1-x  \nonumber
\end{eqnarray}

\noindent These sum rules are the same as those corresponding to the usual
Hubbard model. However, using Eqs.(6) and (7) the following sum rules for
the Cu orbital $d_{x^{2}-y^{2}}$ and "antibonding" O Wannier function $\alpha
$ are obtained: 
\begin{eqnarray}
N^{-}_{d} & = &\langle n_{d\uparrow}+n_{d\downarrow}\rangle = (1+x)
D_{1}^{2} + 2d ( D^{2}_{2} - D^{2}_{1} ) \\
N^{+}_{d} & = & 2(d-x) D^{2}_{1} + (1+x-2d)D^{2}_{2}  \nonumber \\
N^{-}_{p} & = &\langle n_{\alpha\uparrow}+n_{\alpha\downarrow}\rangle =
(1+x) P_{1}^{2} + 2d ( P^{2}_{2} - P^{2}_{1} )  \nonumber \\
N^{+}_{p} & = & 2(d-x) P^{2}_{1} + (1+x-2d)P^{2}_{2}  \nonumber
\end{eqnarray}

\noindent where $d=<n_{i\uparrow}n_{i\downarrow}>$. Adding the four
functions, the total spectral weight becomes: 
\begin{eqnarray}
N_{t}=(1+x)(1+D_{2}^{2}+P_{2}^{2}) - 2x
\end{eqnarray}
\noindent Since $D_{2}^{2}+P_{2}^{2} < 1$, then $N_{t}< 2$. The rest of the
spectral weight has been projected out of the low-energy effective extended
Hubbard model and it is contained at higher energies. The spectral weights
given by Eqs.(19) as a function of doping are represented in Fig.6 for a
typical set of parameters of the three-band model. In Fig.7, we show the
coefficients $P_{i}~$,$~D_{i}$ of the effective operators ( Eq.(8) ) as a
function of the three-band model parameter $\Delta$. We also show the
function $[1-(D_{2}^{2}+P_{2}^{2})]$, related by Eq.(20) to the amount of
spectral weight projected out of the effective low-energy Hamiltonian. From
the slopes of $N^{-}_{d}$ and $N^{-}_{p}$ as a function of doping ( Fig.6 )
one realizes that when doping the stoichiometric compound with electrons,
the latter occupy mainly Cu states, while for hole doping, holes occupy
mainly O states, and for the parameters of Fig.6, part of the Cu holes are
transferred to O holes. This is a well known effect of the Cu-O repulsion $%
U_{pd}$\cite{sim1}.

From the above discussion, it is clear that the main features ( peaks and
valleys) of the Cu and O spectral densities are present in the corresponding
extended Hubbard result ( obtained with the $c_k$ and $c^{\dagger}_k$
operators ), and as a first approximation, the former can be obtained from
the latter modifying the spectral weights of the lower and upper Hubbard
bands with the coefficients given by Eqs.(8) and represented in Fig.7. Thus
in the following we discuss the general properties of the extended Hubbard
spectral density ( independently of the mapping procedure ) and the spectral
weight of both extended Hubbard bands.

In Fig.8 we show the evolution with hole doping of the spectral density of
states of the extended Hubbard model. As discussed previously in the case of
the ordinary Hubbard model\cite{dag2}, after doping, states appear in the
gap of the stoichiometric compound and the Fermi level jumps to the upper
Hubbard band. According to our results, these states are distributed evenly
inside the gap and there is not an impurity band at a defined energy. Also,
the pseudogap persits for all dopings, and both bands evolve smoothly with
doping, with a noticeable transfer of spectral weight from the lower to the
upper band. In Fig.9, we show the same results for the set of parameters 3
of Tables I and II, for which the ratio of the on-site Coulomb repulsion to
hopping parameters is larger. In this case, the weight of the states which
appear in the gap after doping is much smaller, and also the amount of
spectral weight transfer between the bands is smaller. Note also and by
contrast to the usual Hubbard model the strong asymmetry of $\rho(\omega)$
at $x=0$ due to $t_{AA}\neq t_{BB}$.

The transfer of spectral weight in the Hubbard model has been discussed
previously \cite{hybe,eske} and is important in the analysis of x-ray
absorption spectra in the cuprates\cite{chen,hybe}. In the strong-coupling
limit ( large $U$ ), or in the Hubbard III approximation, a static spectral
weight of 2$x$ is easily obtained ( $1-x$ states lie in the upper Hubbard
band). However, as the hopping increases, a positive dynamical contribution
to the spectral weight becomes important. In Fig.10 we show this dynamical
contribution for the extended Hubbard model and the three sets of parameters
of Table II. It is apparent that for realistic parameters for the cuprates,
the dynamical contribution is very important, and as could be seen comparing
Figs. 8 and 9, it is more important for smaller ratios $U/t_m$, where $%
t_{m}=( t_{AA}+2t_{AB}+t_{BB})/4$ is the average effective hopping.

\section{\bf The optical conductivity}

For the evaluation of Eqs.(10) and (11), we have taken the distance between
planes $d_{\perp}=6.64\AA$, corresponding to La$_2$CuO$_4$. Thus, the Drude
weight $D$ and the regular part of the optical conductivity $%
\sigma_{reg}(\omega)$ are proportional to the constant $2e^{2}/(\hbar
d_{\perp})=3.7\times 10^{3}\Omega^{-1}cm^{-1}$. The resulting Drude weight
and average value of the kinetic energy as a function of hole doping are
represented in Fig.11 for the choice of parameters 3 of Table II. The
corresponding results for the optical conductivity are shown in Fig.12. The
average over all boundary conditions has been replaced by a discrete sum
over a square mesh of 8x8 points ($\phi_{1},\phi_{2}$) ( 34 non-equivalent
by symmetry ). For more than 16 points the results are practically
independent on the number of different boundary conditions taken.

For the semiconducting system $(x=0)$, there is a spurious peak at $%
\omega\sim 0.4$. This is a finite-size effect related with the fact
discussed in section III, that the ground state for general boundary
conditions carry a current for finite $U$. The peak corresponds to
transitions to excited states which differ from the ground state mainly in
the spin arrangement. For realistic and large $U$ the magnitude of the
spurious peak is small. In Fig.13, we show the optical conductivity for a
smaller value of the effective $U$ after averaging over 32 ($%
\phi_{1},\phi_{2}$) points ( 18 non-equivalent by symmetry). The magnitude
of the spurious peak increases, but all the other features are very similar
between them, and qualitatively similar to previous results obtained in
periodic 4x4 clusters\cite{dag1}. However, our procedure of averaging
(Eq.(11) ) over boundary conditions leads to more reasonable ( small and
positive ) values of the Drude weight for the insulating system: $%
5\times10^{-5}$ and $3\times10^{-3}\Omega^{-1}cm^{-1}$ for the set 3 and 2
of parameters respectively.

For $x\leq 0.2$, the results are in semiqualitative agreement with
experiments \cite{uchi,falck}. Excluding the spurious peak near $0.4$, the
optical conductivity of the insulating compound shows a gap of $\sim 2$eV
and a characteristic shape at larger energies. Moreover, if the scale of
energies is modified by a $\sim 3/4$, the main features of Fig.13 for $x=0$
coincide with the observed spectral dependence of the persistent
photoconductivity in YBa$_{2}$Cu$_{3}$O$_{6.38}$\cite{ayac}. As the system
is doped the spectral intensity above $\sim 2$eV decreases and at the same
time, a mid-infrared peak appears. Although this low-frequency peak has been
discussed before, its nature has not been clarified yet. Recently, an
explanation based on the string picture of the $t-J$ model has been proposed 
\cite{eder}. In order to shed light on the origin of the different
contributions to the optical conductivity, we have separated the
contributions of the three currents, $j_{AA},~~j_{BB}$ and $j_{AB}$, and 
compared them
with the spectral density, as shown in Fig.14. For hole-doped systems, the
contribution of $j_{AA}$ is negligible. From the analysis of Fig.14, we
conclude that the optical conductivity at energies of the order of the gap
and above corresponds to transitions between the lower and upper extended
Hubbard bands, originated by $t_{AB}$ ( which corresponds to a hopping
between two nearest-neighbor singly-occupied sites ). Instead, the
mid-infrared peak corresponds to one-particle excitations inside the upper
Hubbard band ( from below to above the Fermi level ), originated by $t_{BB}$
or in other words ( in agreement with Ref.\cite{eder} ) to movements of the
added holes without changing the amount of doubly occupied sites. The
difference between the total optical conductivity and the contributions of $%
j_{AB}$ and $j_{BB}$ represented in Fig.14 is mainly due to cross terms
involving both currents.

In Fig.15 we show the frequency dependent effective number of carriers
defined by 
\begin{eqnarray}
N_{ef}(\omega) = {\frac{ {2m_{0}V}}{{\pi e^{2}N} }} \int_{0}^{\omega}
\sigma(\omega^{\prime}) d\omega^{\prime},
\end{eqnarray}

\noindent for the two sets of parameter used before in this section. In
order to compare with experiment, we took the lattice parameter of the CuO$_2
$ planes $a=3.78\AA$, which corresponds to La$_2$CuO$_4$. As well as the
results shown in Figs. 12 and 13, the agreement with experiment is good for $%
x\leq 0.2$. Experimentally, for $x\sim 0.3$ there seems to be an abrupt
change of regime to an uncorrelated metal, with the Drude peak as the only
significant feature of the optical conductivity. Instead, our results do not
show such a transition, but only a smooth evolution. Possible reasons of
this discrepancy are discussed in the next section.

\section{\bf Summary and discussion }

Using an effective extended Hubbard model for the cuprates, we have
calculated the Cu and O spectral density and optical conductivity in a
square cluster of 10 unit cells, averaging over all possible boundary
conditions. To our knowledge, this is the larger cluster for which Cu and O
spectral densities are calculated using the Lanczos method. The averaging
procedure allows to obtain continuous spectral densities and optical
conductivity, and leads to more reasonable values of the Drude weight near
stoichiometry.

We obtain that the states which appear in the gap after doping are
distributed evenly inside it, without building a new band of midgap or
impurity like states. We also obtain that a marked pseudogap persists for
all dopings, separating two bands. When the insulator is doped, the Fermi
level jumps into the band which corresponds according to the sign of the
doping. As doping proceeds, spectral weight is transferred to the band which
contains the Fermi level from the other one. An analysis of the different
components of the current operator allows us to conclude that the
mid-infrared peak in the optical conductivity observed in doped systems is
related to intraband transitions across the Fermi level. The shape of the Cu
and O spectral densities is consistent with the fact that for hole doping,
holes enter mainly O $p_\sigma$ states, while Cu states are occupied on
electron doping.

In general, these results are in agreement with experiment. There is however
a disagreement with some optical experiments\cite{romb,alle} which indicated
almost no shift of the Fermi level with doping, as mentioned in section I.
We should mention that, as discussed by Hybertsen {\it et al.}\cite{hybe}
the substitution of La$^{+3}$ by Sr$^{+2}$ creates a potential which might
localize the doped holes by small doping, creating an impurity band of
acceptor levels like in ordinary semiconductors. To include these effects it
would be necessary to generalize the three-band Hubbard model and the
effective models derived from it.

We do not obtain a deep reconstruction of the electronic structure around
hole doping for $x\sim 0.3$ as conductivity\cite{uchi}, Hall and other
experiments\cite{sree} suggest. A closing of the Mott-Hubbard gap was
expected for similar dopings\cite{sree}. This closing in fact can occur in
the three-band Hubbard model and the effective extended one band model for $%
U_{pd}\sim 3-4$. For these large values of $U_{pd}$ the effective $U$ of the
one band model decreases with doping and a large transfer of holes from Cu
to O takes place \cite{sim1}. However, these values of $U_{pd}$ are too
large compared to the most accepted ones. For realistic parameters of the
three-band model, the effective model parameters are independent of doping 
\cite{sim1}. This does not exclude the possibility that the three-band
parameters change near $x\sim 0.3$, as a consequence of screening caused by
the added holes, for example. Nevertheless, the fact that there are always
two bands as a function of doping is in agreement with x-ray absorption
experiments in La$_{2-x}$Sr$_x$CuO$_4$\cite{chen,hybe}.

Due to the small size of the cluster, we have looked neither for the
presence of a Kondo-like peak near the Fermi energy\cite{roze}, nor the
effects of excitons in the optical conductivity, which are present in the
effective model for realistic and large $V_{ij}$\cite{rap}. A signal of the
presence of these excitons is a smaller gap in the optical conductivity than
in the spectral density. For example, for zero hoppings, it is easy to see
that the gap in the spectral density is given by 
\begin{eqnarray}
E_{gap} & = & U+8(V_{12}-V_{11})
\end{eqnarray}

\noindent However, if the hole and doubly occupied sites are near each
other, the excitation energy is $2V_{12}-V_{11}$ smaller.

We conclude that for the most interesting doping levels and the most
accepted values of the parameters of the three-band model, the general
features of the spectral properties of this model and of the cuprates, can
be described well using an effective extended one-band Hubbard model.

\section{\bf Acknowledgements }

We acknowledge useful conversations with J. Lorenzana. M.E.S. and E.R.G. are
supported by CONICET, Argentina. A.A.A. is partially supported by CONICET,
Argentina.

\newpage

\figure {\noindent FIGURE 1:\\\noindent Scheme of the supercell ( cluster )
containing 10 unit cells used in our calculations. The vectors ${\bf V}_1$
and ${\bf V}_2$ generate the whole lattice by succesive translations of the
supercell.}

\figure {\noindent FIGURE 2:\\\noindent Average optical conductivity as a
function of frequency for Hubbard rings of different length and $U/t=4$.}

\figure {\noindent FIGURE 3:\\\noindent Same as Fig.2, for $U/t=6$.}

\figure {\noindent FIGURE 4:\\\noindent Same as Fig.2, for $U/t=8$.}

\figure {\noindent FIGURE 5:\\\noindent Inverse photoemission ($\omega <
\epsilon_{F}$) and photoemission ($\omega > \epsilon_{F}$) hole spectral
densities for Cu ( dashed line), O ( full line ) and Hubbard operator (
dotted-dash line ), for different hole dopings $x$. The vertical line
indicates the position of the Fermi level $\epsilon_F$. Parameters are given
by the set 1 of table I. The average over boundary conditions have been
replaced by a discrete sum over 4x4 ($\phi_{1},\phi_{2}$) points ( 6
independent points).}

\figure {\noindent FIGURE 6:\\\noindent Total photoemission ($N^+$) and
inverse photoemission ($N^-$) spectral weight for Cu ( $N_d$ ) and O ($N_p$)
as a function of hole doping. Parameters are $U_{d}=9$,$\Delta=3$,$U_{pd}=1$,%
$t_{pp}=0.5$,$U_{p}=4$.}

\figure {\noindent FIGURE 7:\\\noindent Parameters which define the
effective low-energy Cu and O operators according to Eq.(8) as a function of 
$\Delta$. Also shown is $1-D^{2}_{2}-P^{2}_{2}$, which is related with the
high-energy spectral weight projected out of the low-energy effective
Hamiltonian ( see Eq.(20)). Other parameters as in Fig.6.}

\figure {\noindent FIGURE 8:\\\noindent Extended Hubbard inverse
photoemission ($\omega < \epsilon_{F}$) and photoemission ($\omega >
\epsilon_{F}$) hole spectral densities for different hole dopings $x$. The
vertical line indicates the position of the Fermi level $\epsilon_{F}$. 
Parameters are given by the set 1 of table II.}

\figure {\noindent FIGURE 9:\\\noindent Same as Fig.8 for the set 3 of table
II.}

\figure {\noindent FIGURE 10:\\\noindent Total amount of occupied hole
states in the upper band $W$ minus two times the hole doping $x$, as a
function of $x$ for the three sets of parameters of Table I and II. For $%
U\rightarrow\infty$, $W-2x = 0 $.}

\figure {\noindent FIGURE 11:\\\noindent Drude weight and average kinetic
energy of the effective model Eq.(5) as a function of doping. Parameters are
given by the set 3 of table II.}

\figure {\noindent FIGURE 12:\\\noindent Optical conductivity for different
doping levels and the set 3 of parameters of table II. The delta function
contribution at $\omega=0$ has been replaced by a Lorentzian of with 0.01 to
facilitate the comparison with experiment.}

\figure {\noindent FIGURE 13:\\\noindent Same as Fig.12 for the set 2.}

\figure {\noindent FIGURE 14:\\\noindent Bottom: spectral densities of $%
Imag<<j_{i}|j_{i}>>$ ( contributions to the optical conductivity ) of the
partial currents $j_{AB},~~j_{BB}$ and the total result ( see Eqs.(10) and
(13) ) for $x=0.2$ and the set 2 of parameters of Table II. Top:
corresponding hole spectral density. The arrows indicate the transitions
which give rise to the different contributions to the optical conductivity.}

\figure {\noindent FIGURE 15:\\\noindent Effective number of carriers
defined by Eq.(21) as a function of frequency for different doping levels
and the sets of parameters 3 and 2 of Table II.}

\newpage

\centerline{ TABLES }

{\noindent TABLE I : Different sets of parameters of the three-band Hubbard
model $H_{3b}$ ( Eq.1) in units of $t_{pd}$.}\\

\begin{center}
$$
\begin{array}{||c|c|c|c|c|c||}
\hline\hline
Set & \Delta & U_{d} & U_{pd} & t_{pp} & U_{p} \\ \hline
1 & 2.0 & 7 & 1 & 0.5 & 4 \\ \hline
2 & 2.5 & 9 & 1 & 0.5 & 4 \\ \hline
3 & 4.0 & 7 & 1 & 0.5 & 4 \\ \hline\hline
\end{array}
$$
\end{center}

\vskip 1.5 truecm {\noindent TABLE II: Effective one-band parameters for the
three sets of parameters of $H_{3b}$ listed in Table I. The last column
gives the ratio of the effective on-site repulsion $U$ to the average
effective hopping $t_{m}= ( t_{AA}+2t_{AB}+t_{BB})/4$.} \\

\begin{center}
$$
\begin{array}{||c|c|c|c|c|c|c|c|c||}
\hline\hline
Set & U & t_{AA} & t_{AB} & t_{BB} & V_{11} & V_{12} & V_{22} & U/t_m \\ 
\hline
1 & 2.15 & -0.382 & -0.385 & -0.337 & 0.123 & 0.170 & 0.178 & 5.78 \\ \hline
2 & 3.01 & -0.377 & -0.324 & -0.338 & 0.119 & 0.162 & 0.181 & 8.85 \\ \hline
3 & 3.37 & -0.268 & -0.327 & -0.337 & 0.066 & 0.125 & 0.156 & 10.71 \\ 
\hline\hline
\end{array}
$$
\end{center}

\vskip 1.5 truecm {\noindent TABLE III: 
Exact gap $E_{g}$ compared with the gap $E_{g}^{av}$ obtained after averaging
over boundary conditions and $E_{g}^{f}$ for fixed ( antiperiodic ) boundary
condition for a ring of 8 sites and different values of $U/t$.} \\

\begin{center}
$$
\begin{array}{||c|c|c|c||}
\hline\hline
U/t & E_{g} & E_{g}^{av}(8) & E_{g}^{f}(8) \\ \hline
4 &   1.2868  & 1.98 &  3.68\\ \hline
6 &   2.8928  & 3.50 &  4.90\\ \hline
8 &   4.6796  & 5.21 &  6.47\\ \hline\hline
\end{array}
$$
\end{center}


\begin{references}
\bibitem{dess}  D. S. Dessau, Z-X Shen, D. M. King, D. S Marshall, L. W.
Lombardo, P. H. Dickinson, A. G. Loeser, J. Dicarlo, C-H Prak, A.
Kapitulnik, and W. E. Spicer, Phys. Rev. Lett. {\bf 71}, 2781 (1992).

\bibitem{sree}  K. Sreedhar and P. Ganguly, Phys. Rev. B {\bf 41}, 371
(1990).

\bibitem{uchi}  S. Uchida, T. Ido, H. Takagi, T. Arima, Y. Tokura y S.
Tajima, Phys. Rev. B {\bf 43}, 7942 (1991).

\bibitem{falck}  {\ J.P.Falck, A Levy, M. Kastner and R. Birgeneau, Phys.
Rev. Lett. {\bf 69}, 1109 (1992); J. Falck, J. Perkins, A. Levy, M. Kastner,
J. Graybeal and R. Birgeneau, Proc. 2nd Workshop on Phase separation in
Cuprate Superconductors, Cottbus-Germany, Springer-Verlag (1993); Mc Bride,
Miller and Weber, Phys. Rev. B {\bf 49}, 12224 (1994).}

\bibitem{nuck}  N. N\"{u}cker in {\it Physics of High-Temperature
superconductors}, edited by S. Maekawa and M. Sato (Springer Verlag, Berlin,
1991), p. 283.

\bibitem{chen}  C.T. Chen, F. Sette, Y. Ma, M. S. Hybertsen, E. B. Stechel,
W. Foulkes, M. Schl\"uter, S. W Cheong, A. S. Cooper, L. W. Rupp Jr.,B
Batlogg, Y. L. Soo, Z. H. Ming, A. Krol, and Y. H. Kao, Phys. Rev. Lett. 
{\bf 66}, 104 (1991).

\bibitem{hybe}  M. Hybertsen, E. Stechel, W. Foulkes, and M. Schl\"{u}ter,
Phys. Rev. B {\bf 45}, 10032 (1992).Phys. Rev. B {\bf 44 }, 7504 (1991).

\bibitem{romb}  H. Romberg, M. Alexander, N. N\"{u}cker, P. Adelmann, and J.
Fink, Phys. Rev. B {\bf 42}, 8768 (1990).

\bibitem{alle}  J. Allen, C. Olson, M. Maple, J. Kang, L. Liu, J. Park, R.
Anderson, W. Ellis, J. Markert, Y. Dalichaouch, and R. Liu, Phys. Rev. Lett. 
{\bf 64}, 595 (1990).

\bibitem{shen}  Z-X Shen, D. S Dessau, B. O. Wells, C. G. Olson, D. B.
Mitzi, L. Lombardo, R. L. List, and A. J Arko, Phys. Rev. B {\bf 44}, 12098
(1991).

\bibitem{perk}  J.D.Perkins, J. Graybeal, M. Kastner, R. Birgeneau, J.
Falck, and M. Greven, Phys. Rev. Lett. {\bf 71}, 1621 (1993).

\bibitem{ayac}  C. Ayache, I. L. Chaplygin, A. I. Kirilyak, N. M. Kreines,
and V. I. Kudinov, Solid State Commun.{\bf 81}, 41 (1992).

\bibitem{tera}  I. Terasaki, T. Nakahashi, A. Maeda, and K. Uchinokura, 
Phys. Rev. B {\bf 43}, 551 (1991) and references therein.

\bibitem{fuji}  {A. Fujimori, I. Hase, M. Nakamura, H. Namatame, Y.
Fujishima, Y. Tokura, M. Abbate, F. M. F. Groot, M. T. Czyzyk, J. C. Fuggle,
O. Strebel, F. L\'opez, M. Domke, and G.Kaindl, Phys. Rev. B {\bf 46}, 9841
(1992).}

\bibitem{roze}  {\ M. J. Rozemberg, G. Kotliar, H. Kajueter, G. A. Thomas,
D. H. Rapkine, J. M. Honig, and P. Metcalf, Phys. Rev. Lett. {\bf 75}, 105
(1995).}

\bibitem{mill}  A. J. Millis and S. N. Coppersmith, Phys. Rev B {\bf 42},
10807 (1990).

\bibitem{wagn}  J. Wagner, W. Hanke and D. Scalapino, Phys. Rev. B {\bf 43},
10517(1991).

\bibitem{esk1}  H.Eskes, M.Meinders and G.Sawatzky, Phys. Rev. Lett.{\bf 67}%
, 1035(1991).

\bibitem{sf}  C. D. Batista and A. A. Aligia, Phys. Rev. B {\bf 47}, 8929
(1993).

\bibitem{che2}  C. Chen, H. Sch\"{u}ttler, and A. Fedro, Phys. Rev. B {\bf %
41 }, 2581 (1990); Physica B {\bf 143}, 530 (1990); C. Chen and H.
Sch\"{u}ttler, Phys. Rev. B {\bf 43}, 3771 (1991).

\bibitem{dag1}  E. Dagotto, R. Joint, A Moreo, S. Bacci and E. Gagliano,
Phys. Rev. B {\bf 41}, 9049 (1990); E. Dagotto, A. Moreo, F. Ortolani, J.
Riera and D. Scalapino, Phys. Rev. Lett. {\bf 67}, 1918 (1991).

\bibitem{dag2}  E. Dagotto, A. Moreo, F. Ortolani, D. Poilblanc and J.
Riera, Phys. Rev. B {\bf 45}, 10741 (1992).

\bibitem{ohta}  Y. Otha, T. Tsuitsui, W. Koshibae, T. Shimozato,  and S.
Maekawa, Phys. Rev. B {\bf 46}, 14022 (1992).

\bibitem{poil}  D. Poilblanc, Phys. Rev. B {\bf 44}, 9562 (1991).

\bibitem{lore}  J. Lorenzana and L Yu, Phys. Rev. Lett. {\bf 70}, 861 (1993).

\bibitem{hors}  P.Horsch and W.Stephan, Phys. Rev. B{\bf 48},10595(1993).

\bibitem{bulu}  N. Bulut, D. J. Scalapino, and S. R. White,  Phys. Rev.
Lett. {\bf 72}, 705 (1994).

\bibitem{jakl}  J. Jaklic and Prelovsek, Phys. Rev. Lett. {\bf 75}, 1340
(1995).

\bibitem{eder}  R. Eder, P. Wrobel, and Ohta,Phys. Rev. B{\bf 59}, 11034
(1996).

\bibitem{eske}  {H. Eskes and A. Ole\'s, Phys. Rev. Lett. {\bf 73}, 1279
(1994); H. Eskes, A. Ole\'s, M. B. Meinders, and W. Stephan, Phys. Rev. B 
{\bf 50}, 17980 (1995).}

\bibitem{unge}  P. Unger and P. Fulde, Phys. Rev. B {\bf 51}, 9245 (1995).

\bibitem{chub}  A. V. Chubukov, D. K. Morr, and K. A. Shakhnovich,  
{\it preprint}.

\bibitem{3b}  V.J. Emery, Phys. Rev. Lett. {\bf 58}, 2794 (1987); P.B.
Littlewood, C.M. Varma, and E. Abrahams, Phys. Rev. Lett. {\bf 63}, 2602
(1989); references therein.

\bibitem{fei1}  L.F. Feiner, J. H. Jefferson, and R. Raimondi, Phys. Rev. B 
{\bf 53}, 8751 (1996); references therein.

\bibitem{sim1}  M. E. Sim\'on and A. A Aligia, Phys. Rev. B {\bf 53}, 15327
(1996); references therein.

\bibitem{erol}  J. M. Eroles, C. D. Batista, and A. A. Aligia, Physica C 
{\bf 261}, 137 (1996); C. D. Batista and A. A. Aligia, Physica C {\bf 264},
319 (1996); references therein.

\bibitem{cast}  H. Castillo and C. Balseiro, Phys. Rev. Lett. {\bf 68}, 121
(1992).

\bibitem{shastry}  B.S.Shastry, B.I.Shraiman, and R.R.P.Singh, Phys. Rev.
Lett. {\bf 70}, 2004(1993).

\bibitem{rojo}  A. Rojo, G. Kotliar and G. Canright, Phys. Rev. B {\bf 47},
9140 (1993).

\bibitem{assa}  F. F. Assad and M. Imada, Phys. Rev. Lett. {\bf 74}, 3868
(1995).

\bibitem{gros}  C. Gross, Z Phys. B {\bf 86}, 359(1992).

\bibitem{xiang}  T. Xiang and J. M. Wheatly,  Phys. Rev. B{\bf 54}, 
12653(1996).

\bibitem{ovch}  A.A.Ovchinnikov, Soviet Phys. JETP {\bf 30}, 1160(1970) and
references therein.

\bibitem{fei2}  L. Feiner, Phys. Rev. B {\bf 48}, 16857 (1993).

\bibitem{taki}  M. Takigawa, P. C Hammel, R. H. Heffner, Z. Fisk, K. C. Ott,
and J, D, Thompson, Phys. Rev. Lett. {\bf 63},1865 (1989).

\bibitem{pele}  J. Tranquada, S. Heald, A. Moodenbaugh, and M. Suenaga,
Phys. Rev B {\bf 35}, 7187 (1987); A. Fujimori, E. Takayama-Muromachi, Y.
Uchida, and B. Okai, Phys. Rev. B {\bf 35}, 8814 (1987); E. Pelegrin, N.
N\"{u}cker, J. Fink, S. L. Molodtsov, A. Guti\'{e}rrez, E. Navas, O.
Strebel, Z. Hu, M. Domke, G. Kaindl, S, Ushida, Y. Nakamura, J. Markl, M.
Klauda, G. Saemann-Ischenko, A. Krol, J. L. Peng, Z. Y. Li, and R. L.
Greene, Phys. Rev. B {\bf 47}, 3354 (1993).

\bibitem{anne}  J.F.Annett, R.M.Martin, A.K.McMahan, and S.Sarpathy, Phys.
Rev. B{\bf 40}, 2620(1989); J.B. Grant and A.K. McMahan, Phys. Rev. Lett. 
{\bf 66},488 (1991).

\bibitem{hyb2}  M.S. Hybertsen, E.B. Stechel, M. Schl\"{u}ter and D.R.
Jennison, Phys. Rev. B {\bf 41}, 11068 (1990).

\bibitem{liu}  R.Liu, D. Salamon, M. Klein, S. Cooper, W. Lee, S-W Cheong,
and D. Ginsberg, Phys. Rev. Lett. {\bf 71}, 3709 (1993); D. Salamon, R. Liu,
M. Klein, M. Karlow, S. Cooper, S-W Cheong, W. Lee, and D. Ginsberg, Phys.
Rev. B{\bf 51}, 6617 (1995).

\bibitem{rap}  M. E. Sim\'on, A. A. Aligia, C. D. Batista, E. R. Gagliano,
and F. Lema, Phys. Rev. B {\bf 54}, R3780 (1996).


\bibitem{schu}  H.B. Sch\"{u}ttler and A.J. Fedro, Phys. Rev. B {\bf 45},
7588 (1992).

\bibitem{hirs}  J. Hirsch, Physica C {\bf 158}, 326 (1989); Phys. Lett. A 
{\bf 134}, 351 (1989); F. Marsiglio and J. Hirsch, Phys. Rev. B {\bf 41},
6435 (1990).

\bibitem{japa}  Japaridze and E. M\"uller-Hartmann, Ann. Phys. (Leipzig) 
{\bf 3}, 163 (1994).

\bibitem{arr1}  L. Arrachea, A. A. Aligia, E. R. Gagliano, K. Hallberg, and
C. A. Balseiro, Phys. Rev. B{\bf 50}, 16044 (1994).

\bibitem{mich}  K. Michelsen and H. De Raedt, Int. J. Mod. Phys. {\it to be
published}.

\bibitem{arr2}  L. Arrachea, A. A. Aligia, and E. R. Gagliano, Phys. Rev.
Lett. {\bf 76}, 4396 (1996); references therein.

\bibitem{ali1}  A. A. Aligia, L. Arrachea, and E. R. Gagliano, Phys. Rev. B 
{\bf 51}, 13774 (1995); references therein.

\bibitem{oht2}  Y. Otha, T. Tsuitsui, W. Koshibae, and S. Maekawa, Phys.
Rev. B {\bf 50}, 13594 (1994).

\bibitem{gag1}  E. R. Gagliano, A. A. Aligia, L. Arrachea, and M. Avignon, 
Phys. Rev. B {\bf 51}, 14012 (1995).

\bibitem{trip}  M. E. Sim\'on and A. A. Aligia, Phys. Rev. B {\bf 52}, 7701
(1995).

\bibitem{ali2}  A. A. Aligia, M. E. Sim\'on and C. D. Batista, Phys. Rev. B 
{\bf 49}, 13061(1994).

\bibitem{gag2}  E. Gagliano and C. Balseiro, Phys. Rev. Lett. {\bf 59}, 2999
(1987).

\bibitem{blou}  E. I. Blount, Phys. Rev. B {\bf 38}, 6711(1988).

\bibitem{sim3}  M. E. Sim\'on and A. A. Aligia, Phys. Rev. B {\bf 46}, 3676
(1992).

\end{references}
\end{document}